\documentclass[showpacs,aps,superscriptaddress,prd,notitlepage,showkeys,nofootinbib]{revtex4-1}
\usepackage[normalem]{ulem}
\usepackage{multirow}
\usepackage{amssymb}
\usepackage{amsmath}
\usepackage{graphicx}
\usepackage{dcolumn}
\usepackage{hyperref}

\hypersetup{colorlinks,urlcolor=cyan,citecolor=cyan,linkcolor=cyan}
\usepackage{color,units}
\usepackage[dvipsnames]{xcolor} % for more colours!
\usepackage{lineno}
\usepackage{xspace}
\usepackage{longtable} 
\usepackage{float}  
\usepackage{aas_macros}
\usepackage{amsfonts,wasysym,epsfig,verbatim,subfigure,bm,mathrsfs,lipsum}

\usepackage{orcidlink}

\usepackage{color}
\usepackage{hyperref}
\usepackage{amsmath}
\usepackage{multirow}
\usepackage{graphicx}
\usepackage{envmath}
\usepackage{natbib}

\newcommand{\WSU}{Department of Physics \& Astronomy, Washington State University, 1245 Webster, Pullman, WA 99164-2814, USA} 

\begin{document}

\title{\texttt{CoBiTS}: Single-detector discrimination of binary black hole signals from glitches using deep learning}
\author{Matthew VanDyke, Kexuan Wu, and Sukanta Bose}
%\author{Matthew VanDyke\corref{}}
%\author{Kexuan Wu\corref{}}
%\author{Sukanta Bose\corref{}}
%\orcidlink{0000-0002-4151-1347}
%\address{\WSU}
\affiliation{\WSU}
%\author{Sukanta Bose\corref{cor1}}
%\ead{sukanta@wsu.edu}

\begin{abstract}
We develop a Conformer neural network, called Conformer Binary
neTwork Search, or \texttt{CoBiTS}, for distinguishing binary black
hole gravitational wave (GW) signals from non-Gaussian and
non-stationary noise artifacts in the data from current generation
LIGO-Virgo-KAGRA detectors. A large subset of these transient noise
artifacts, termed as ``glitches'' for short, trigger BBH search
templates. Some of them go on to produce detection candidates
and require human vetting, supported by data quality tools, to be
correctly identified and vetoed. In its current version, \texttt{CoBiTS} takes
as inputs single-detector strain timeseries snippets, claimed by other search
pipelines to be containing GW candidates, and outputs the 
significance of each snippet to contain a BBH signal and a glitch. 
\texttt{CoBiTS} is shown to be particularly effective in discriminating
high-mass BBH signals from blips and scattered light glitches, even
when a signal is near concurrent or overlapping with a glitch. 
The performance of \texttt{CoBiTS} gains from employing Conformer, which is a
specialized model that combines convolutional layers and Transformer
architecture for sequence modeling tasks. Conformer is especially good at  
leveraging the strengths of both convolutional layers -- for local
feature extraction -- and self-attention layers -- for capturing
long-range dependencies. 
\end{abstract}

\maketitle

\section{Introduction}
\label{sec:intro}

The launch of gravitational wave (GW) astronomy has had a stellar
start~\cite{GW150914,TheLIGOScientific:2017qsa,LIGOScientific:2025slb},
with all sources detected so far as binaries involving stellar-mass or intermediate-mass black holes and neutron stars. This work develops solutions that will provide better
quality alerts to external observatories and detectors -- even when
only one GW interferometer registers such a binary as a detection candidate. 
Since we expect the CBC event rate to be considerably higher in the fifth observation run (O5) than in past runs of current ground based detectors, it assumes greater significance that we step up the GW alert “quality control” or even automate it (see Ref.~\cite{Essick:2020qpo} as an example of such an exercise).
This is why we propose here a Machine Learning (ML) algorithm for better and faster discrimination of binary black hole (BBH) signals from noise transients~\cite{LIGO:2024kkz} 
than traditional methods, in data from a single detector. Specifically, we 
develop a high-mass
BBH candidate vetting pipeline that: 
(a) is  computationally faster than matched filtering; 
(b) is at least as good as matched filtering, in its volume sensitivity, to BBHs across luminosity distance;
(c) performs better than
matched filtering, especially, when non-stationary or non-Gaussian noise transients, or ``glitches'', are  overlapping or present  within a chirp-time of  a BBH signal. This is observed for a 
large subset of the non-spinning BBH parameter space. 

The reason we focus on improving the vetting of single-detector search candidates is multifold. First, it is the basic unit of a transient search with a network of detectors. Second, since the duty factor is not going to be 100\% in the current generation kilometer-scale detectors, improving the ability to detect with single detectors will improve the overall BBH detection rates of an observation run with any network. Third, owing to varied sensitivities and the different orientations of the global detectors,~\footnote{LIGO Hanford and Livingston detectors have very similar but not exactly the same antenna patterns.}
occasionally, signals from the same binary can turn out to be above the detection threshold in one detector but not the others.
Further, even a moderately significant signal in one detector can rise in significance when analyzed in conjunction with data from other detector(s) if they were in observation mode but did not on their own find an above-threshold signal around that time. A fast vetting of such a single-detector candidate can be exploited to launch follow-up multi-detector analyses.
Future versions of search pipelines can integrate our network model's functions in order to reduce their false-alarm rates and event retractions.
Traditional BBH search pipelines~\cite{2016_pycbc_usman,aubin_mbta_2021,Sachdev:2019vvd, chu2021spiir} 
have mechanisms for discriminating against the primary glitches~\cite{Nitz_2018,Sachdev:2019vvd}
with varying degrees of effectiveness. However, since their optimal statistics make simplifying assumptions about the detector noise, which may not be an accurate characterization of the real data, the ML models hold promise for better performance. They can be trained to construct better empirical representations of that data.

Almost all BBH search pipelines in ML to-date that satisfy the characteristics (a) and (b) above, in real strain data, have been shown to do so when employing multi-detector coincidence. Our single-detector improvements are promising and hold out hope for further performance upticks that can contribute to increasing the BBH detection rate. This will help in forming a more accurate picture of BBH demographics, which informs our understanding of their formation channels, population synthesis models, and aids broader pursuits like the measurement of the Hubble constant.
The reason it helps to achieve computational speed-up for transient GW searches is that it enables  followup campaigns by other observatories for catching any electromagnetic or particle counterpart signals that may get triggered by their astrophysical sources.
As noted above, an MMA observation of this kind can significantly boost the probability of finding the progenitor's cosmic coordinates or the true host galaxy. This in turn can reveal any peculiarities that maybe shared by those hosts or stellar-mass BBH environments and, thereby, shed light on their formation scenarios. A host galaxy can also yield redshift information, which along with the GW luminosity distance can provide a measurement of the Hubble constant.
On the other hand, any reductions in the
computational costs of signal searches can help in allocating the freed-up resources for other scientific tasks.
The only  
GW signal that is unequivocally accepted to have had  electromagnetic counterparts is the binary neutron start (BNS) merger GW170817~\cite{TheLIGOScientific:2017qsa}. Arguably, BNSs should be the top target of such ML models. This is what we will pursue in the future but restrict ourselves here to BBHs, which have shorter signals. BBHs have also been proposed as progenitors of a few EM signals and, if true, offer intriguing prospects for probing their host environments~\cite{Graham:2020gwr}.

We will target the era of the fifth observation run (O5) of the ground-based GW detectors LIGO~\cite{advligo,Saleem_2022}, Virgo~\cite{Acernese_2015} and KAGRA~\cite{kagra}. 
We expect several GRB, X-ray, optical and radio observatories (not to mention neutrino detectors) to be active then that can contribute to this MMA campaign. New campaigns, such as the Legacy Survey of Space and Time (LSST) of the Rubin Observatory, may also capitalize on MMA opportunities~\cite{LSST:2008ijt}.

In recent years, multiple machine learning (ML) approaches have been proposed 
for gravitational-wave (GW) data analysis (see~\cite{Cuoco_2021} for a review). 
A key motivation is that neural networks can exploit non-linear feature 
relationships to distinguish patterns in time-transient data and images. 
They are particularly effective in learning complex, non-linear structures and 
in separating clusters that are not linearly separable. By contrast, 
traditional matched filtering and signal-based consistency tests, such as 
the $\chi^2$ discriminator~\cite{b_allen,PhysRevD.96.103018}, primarily rely on 
linear decision boundaries to separate compact binary coalescence (CBC) signals 
from noise artifacts.

Many ML models have been developed for compact binary coalescence (CBC) searches~\cite{gabbard_matching_2018,schafer_one_2022,schafer_first_2023,schafer_training_2022,Nousi:2022dwh,wang_gravitational-wave_2020,mishra_optimization_2021,Marx:2024wjt,Nagarajan:2025hws,skliris_real-time_2024}. 
Recent examples include Aframe~\cite{Marx:2024wjt}, which searches directly in GW strain data but requires coincident input from multiple detectors, 
and MLy~\cite{skliris_real-time_2024}, which targets short-duration bursts using a dual-CNN architecture to exploit inter-detector coherence without templates. 
Neither approach is explicitly trained on labeled glitches, and both rely significantly on multi-detector data,
which is a sensible strategy when their observational data are available. 

Several approaches focus on glitch classification using processed representations. 
Glitch is a generic term used to describe transient noise artifacts in gravitational-wave detector data that originate from environmental or instrumental artifacts.
%CITE
Not all glitches have a known origin.
The Gravity Spy project combines multiple machine-learning approaches with citizen science to classify LIGO glitches into morphological categories based on their spectrogram representations~\cite{Zevin_2017, wu2025advancing, soni2021discovering,zevin_gravity_2024}. Magee et al.~\cite{magee2024mitigatingimpactnoisetransients} employ CNNs to classify 
various glitches (e.g., blips and scattering arches) and GW signals from grayscale 
images constructed by applying singular value decomposition (SVD) to signal-to-noise 
ratio (SNR) time transients and transforming into the frequency domain. 
Lopez et al.~\cite{lopez_ameliorating_nodate} combine matched filtering with a 
multi-layer perceptron (MLP) to distinguish intermediate-mass black hole (IMBH) 
signals from glitches, using SNR tracks produced by the matched-filter pipeline as inputs. 
GSpyNetTree~\cite{Alvarez-Lopez:2023dmv} distinguishes signals from glitches 
using time-frequency maps. 
In comparison, \texttt{SiGMa-Net}~\cite{Choudhary:2022yje} targets the same objective but by 
projecting the GW strain data on a basis of sine-Gaussians, parametrized by their central time $t_0$ and quality factor $Q$, and then training their ML model on how differently the projected power is distributed in the $(t_0,Q)$ space for BBH signals and glitches.
All of these methods rely on processed representations rather than raw strain transients.

Waveform reconstruction networks such as AWaRe~\cite{chatterjee_navigating_2024,chatterjee_no_2025}, WaveFormer~\cite{wang_waveformer_2024}, and DeepExtractor~\cite{s91m-c2jw} aim to recover gravitational-wave (GW) signals from noisy time series while accounting for multiple physical parameters. As far as we can assess, neither AWaRe nor WaveFormer was explicitly trained on glitch-contaminated data, although AWaRe maintains reasonable performance in the presence of certain types of glitches~\cite{chatterjee_no_2025}. DeepExtractor, on the other hand, includes attempts to reconstruct simulated glitches.

From ML architecture perspective, ResNet has been adapted in several past ML GW works~\cite{Nousi:2022dwh,Marx:2024wjt,Nagarajan:2025hws}, 
%Aframe:Marx:2024wjt,
%SAGE is Nagarajan:2025hws
and is an effective convolutional neural network mainly used for recognizing images. Thus, ResNet excels when applied to spectrograms of signals and glitches. Indeed, its use of residual connections enables it to train very deep networks efficiently, which has made it a standard tool for computer vision classification and other tasks. While it can be adapted for sequential data, it is not inherently designed for this purpose.

In this paper, we propose a novel ML model designed specifically to distinguish BBH signals from detector glitches using a state-of-the-art architecture. We call our ML model the Conformer Binary neTwork Search, or \texttt{CoBiTS}. From the training and inference perspective, our model operates directly on raw strain data from a single detector and is trained to distinguish binary black hole (BBH) signals from real detector glitches, even when they occur within seconds of each other or overlap in time, and to infer the likelihoods of the existence of both glitches and signals. This enables robust classification without relying on inter-detector coincidence or intermediate feature maps, thereby reducing potential information loss or biases introduced during pre-processing.

From the architecture perspective, unlike any of the ML searches proposed so far for GW signals, our model employs Conformer for the first time. Conformer is a specialized model that combines convolutional layers and Transformer architecture~\footnote{A Transformer model is a type of deep learning architecture that uses the self-attention mechanism to analyze and process sequential data such as text, images, or audio.}, leveraging the strengths of both convolutional layers for local feature extraction and self-attention layers for capturing long-range dependencies. This makes it particularly well-suited for discriminating CBC signals from non-Gaussian and non-stationary noise transients in single-detector strain data.

This paper is organized as follows.
In \hyperref[sec:method]{Section~\ref*{sec:method}}, we describe the methodology, including the machine-learning model architecture and the inference approach.
\hyperref[sec:datasets]{Section~\ref*{sec:datasets}} discusses the simulated and astronomical datasets used in this study.
Our results, including the performance of \texttt{CoBiTS} on real gravitational-wave events, are presented in \hyperref[sec:results]{Section~\ref*{sec:results}}. We conclude with observations on future prospects in \hyperref[sec:conclusions]{Section~\ref*{sec:conclusions}}.

\section{An outline of the classification method}
\label{sec:method}

Matched-filtering noisy data with templates of CBC signals has been the workhorse of CBC searches. Those searches also typically use additional means to discriminate against non-Gaussian or non-stationary noise transients by employing signal or glitch-based discriminators, such as $\chi^2$ tests.
Running these searches and tests in low-latency is important to be able to issue public alerts but is computationally expensive. Deep learning networks can aid by reducing that cost in the execution of the search
while holding promise for somewhat better performance. The improvement in performance can arise from a better modeling of the detector noise than what traditional discriminators or $\chi^2$ tests typically accomplish.

In this work, we develop a machine learning (ML)  pipeline based on {\it Conformer}.
We construct representative datasets, and design and train our network model, aiming for strong generalization performance. This method shifts the majority of the computational complexity to training time. After training, the model can perform searches in real-time on just a single GPU. While matched-filtering typically requires computing  correlations with the same stretch of GW strain data repeatedly with CBC templates looped from a vast template bank, making real-time searches computationally demanding, \texttt{CoBiTS} replaces these operations with a single, efficient forward pass through the GW data, thereby significantly reducing inference time computation.

\subsection{Model Architecture}
\label{subsec:model_archmain}

The model architecture developed for this work is based on a deep Conformer encoder~\cite{gulati2020conformer} tailored to classify GW signals. Since GW strain data closely resembles audio 
timeseries signals, differing primarily in their noise characteristics, we drew inspiration from state-of-the-art audio processing models like Wav2Vec2~\cite{baevskiZhou}.~\footnote{We use a similar architecture, but vastly different training schema. Reference~\cite{baevskiZhou} utilizes self-supervised pretraining, whereas we use fully supervised training. Incorporating semi-supervised pretraining is a future direction we plan to explore with GW data.} 

CNN-based models  
(e.g., Gravity Spy~\cite{Gravityspy2})
inherently have a strong local inductive bias: They excel at identifying compact, local features, which is quite advantageous in searching for CBCs. However, these models can struggle when the data contains extended structures that persist over longer timescales.~\footnote{Note that blip glitches, which are typically very short-lived~\cite{cabero_blip_2019}, alone would likely not benefit from global attention; however, extended noise artifacts such as scattered light~\cite{Soni_2021} do.} In these cases, a broader receptive field becomes the clear solution. When the model can ``see" farther in time, it can incorporate information across a wider interval, recognizing that separate features may come about from the same physical origin (e.g., scattered light). This global view, afforded by self-attention within the Conformer, allows the network to distinguish between extended instrumental artifacts and genuine signals. Thus, the inclusion of attention layers effectively widens the models temporal view.

\begin{figure}
    \centering
    \includegraphics[width=0.7\textwidth]{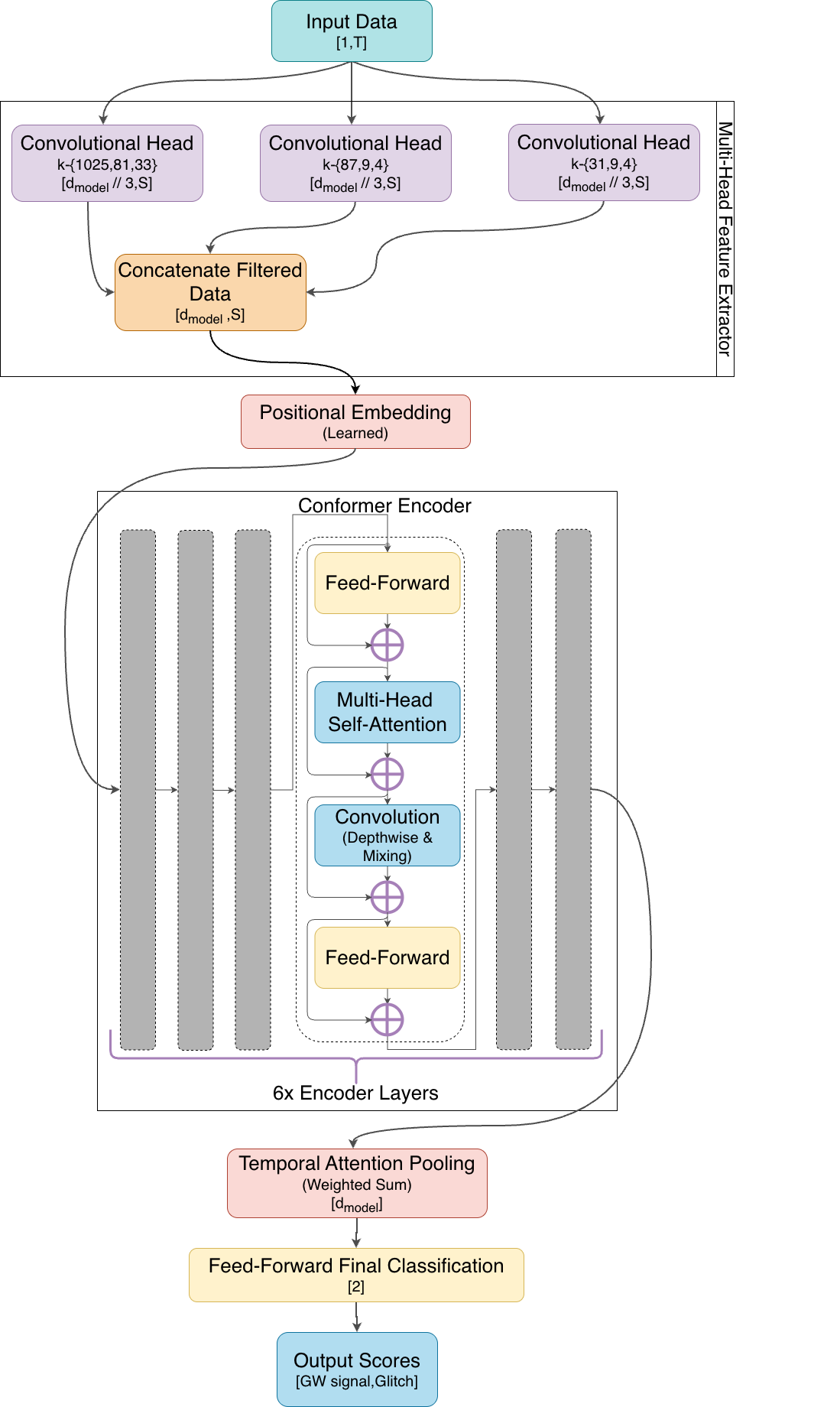}
    \caption{Model diagram/flowchart of \texttt{CoBiTS}. The network inputs timeseries data of a certain duration ($T$) from a single (1) channel, e.g., in the form of a GW strain time-series from LIGO-Hanford. Here,  $[1, \mathrm{T}]$ defines the dimensions of that dataset. That input is passed through the Multi-Head Feature Extractor (MHFA), which comprises three parallel convolutional heads, each consisting of three convolutional layers. The kernel size of each layer is located within the curly braces in the diagram. The first layer of each head expands the input to $[\mathrm{d}_\mathrm{model} / 3, *]$, and the following two layers consist of many-to-many operations preserving this channel dimension. After these convolutional heads, the resultant features are concatenated along the channel dimension to form outputs of shape $[\mathrm{d}_\mathrm{model},\mathrm{S}]$. Second, after a learned positional embedding is added to the data, it passes through six conformer encoder layers. These encoder layers preserve the shape of the data, but process it further while modeling both global and local temporal dependencies. Finally, these attended features are pooled along the temporal dimension, and, through a simple feed-forward dense network, reduced to the two output logits. See Appendix~\ref{subsec:model_arch}.}
    \label{fig:NNworflow}
\end{figure}

As illustrated in Fig.~\ref{fig:NNworflow}, the \texttt{CoBiTS} architecture consists of three primary modules: a Multi-Head Feature Extractor (MHFA), a Conformer Encoder, and a final pooling operation. The MHFA first processes the whitened strain through a set of parallel stride convolutions. These learned filters behave as local feature ``detectors" that both denoise and downsample the input, producing a shorter and more feature-dense representation of the input strain. Conceptually, this stage transforms the time series into a learned time-frequency representation, similar to a spectrogram but optimized for downstream discriminatory tasks.

The \texttt{CoBiTS} architecture employs an initial feature extraction stage for downsampling and denoising the timeseries input data.  followed by a series of Conformer layers, which integrate convolutional operations~\footnote{In machine learning, particularly in CNNs, a convolution operation involves applying multiple learned filters (kernels) across an input 
array, which is a strain timeseries in our case. 
Unlike classical convolution, these operations use multiple learned kernels to transform a single input channel (or many) into many output channels. Some key characteristics of these operations include an adjustable stride (the step size of the kernel as it moves across the input), dilation (the spacing between kernel elements), and padding. These learned filters enable CNNs to identify diverse local features in the input, creating a richer multi-channel representation for subsequent layers.} and self-attention~\footnote{Self-attention is a mechanism that allows a model to weigh the importance of different elements in the input sequence relative to each other. Unlike convolution, self-attention captures global relationships by enabling each element to directly attend to every other element, regardless of distance,
{which in our case is the time-separation between features extracted from the GW timeseries}.} 
mechanisms to capture both local and global context within the data. After an adaptive pooling operation, a classification head maps the learned representations to their appropriate classes. A comprehensive description of each architectural block and the training procedure can be found in Appendix~\ref{subsec:model_arch}.

\subsection{Inference}

For signal inference, the model processes a continuous input stream (timeseries) by sliding a fixed-length window with overlap across the data. Because the model is trained to be time-translationally invariant, true loud signals are expected to flag  multiple, overlapping windows as highly significant to contain a signal. Random fluctuations, by contrast, will rarely coincide in the same way across successive overlapping windows. As a result, true signals appear as ``square-wave" plateaus---elevated scores spanning the full window length---while noise appears as isolated, spurious spikes; see Section~\ref{sec:results} for an example of these contrasting features.
Thus, we improve detection accuracy and suppress noise spikes by filtering the model's output for these ``square wave" patterns using a one-dimensional moving-average filter.

\section{Datasets employed}
\label{sec:datasets}

The datasets utilized in this study comprise real interferometric data from the LIGO O3 data release~\cite{Abbott_2023} and simulated Binary Black Hole (BBH) merger transients, categorized into four primary types: quiet detector noise, transient noise artifacts (glitches), simulated gravitational-wave (GW) signals injected into quiet detector noise, and simulated GW signals injected into transient noise artifacts.
{To promote generalization and reduce training bias, which can arise in limited training data studies, this diverse dataset of signals and detector noise was constructed -- independently for each training epoch.}

\subsection{Detector Noise and Glitches}

In this work, we use real data for both quiet detector noise and glitches. Quiet detector noise segments were sourced from the LIGO Hanford O3 data release~\cite{KAGRA:2023pio}.~\footnote{We run multiple tests in O3 data from LIGO Hanford, as reported below, even though that dataset was not used for training our network model.}
To ensure data purity, segments associated with Gravitational-Wave Candidate Event Database (GraceDB)
triggers were excluded. We obtained almost 26,000
30-second segments with no reported CBC signal or noise transient.
For glitches, we focus on two frequently occurring types that affect the sensitivity of BBH searches the most, namely, the blip and scattered-light glitches. We obtained them  
primarily from the Gravity Spy O3 glitch catalog~\cite{Glanzer:2022avx}, as long as they were classifed as such with no less than 90\% confidence. This selection contained 18,630 glitches, comprising 6,972 blips  and 11,658 scattered-light glitches -- of both slow and fast kinds. 
Data segments containing them were set to match the length of the quiet detector noise segments, each 30-seconds long.

\subsection{Binary Black Hole Signal Simulation and Injection}

Gravitational wave signals from binary black hole (BBH) inspiral-merger-ringdown events were simulated using the \texttt{IMRPhenomXPHM} waveform model~\cite{pratten_computationally_2021}, a frequency-domain model 
of signals from quasi-circular, precessing BBHs across the full inspiral, merger, and ringdown phases. This model incorporates both dominant and sub-dominant harmonic modes, enabling detailed modeling of complex precessing signals. Calibrated against numerical relativity simulations, \texttt{IMRPhenomXPHM} supports mass ratios, $q \equiv {m_2}/{m_1}$, down to approximately $1/18$ and spin magnitudes up to 0.85 with generic spin orientations. In our study, we used this model to generate non-spinning BBH signals with total masses uniformly distributed between 50~$M_\odot$ and 120~$M_\odot$, and the mass ratio ($q$)  uniformly sampled from 0.1 to 1.
The detailed parameter space and discretization used for waveform generation are summarized in Table~\ref{tab:bbh_parameters}. For our injection campaign into real gravitational-wave strain data, this parameter space yielded 10,000 distinct waveforms covering a broad range of physical configurations. We deliberately avoided adopting a BBH mass distribution model—such as the power-law + peak model~\cite{gwtc3} 
because one of our primary goals is to identify regions in the two-dimensional component-mass space where our detection model performs best and worst relative to the traditional matched-filtering pipeline, particularly in the presence of noise transients in real gravitational-wave data.

We adopted a reference low-frequency cutoff of 20Hz.
In the BBH parameter space that we covered in this paper, the longest waveform duration is approximately 5.36 seconds. 
Thus, they fit well within our 
timeseries data that are 22 seconds long, 
which provides ample buffer and the possibility of extending this analysis to lower-mass systems in future studies, including binaries with total masses as low as 25~$M_\odot$ and mass ratios down to 0.1, covering both asymmetric and equal-mass configurations. 
These segments are long enough to include a sufficient duration of quiet detector noise as well as data with the aforementioned  glitches and injected BBH signals.
Only a subset of the slow scatter glitches extended beyond this duration.

To simulate real BBH signals, we employed the
\texttt{IMRPhenomXPHM} waveform and projected it on the detector assuming the
sources to be randomly distributed in the sky as described in  
Table~\ref{tab:bbh_parameters}.
The sources were taken to be uniformly distributed in volume, with the luminosity distance $d_L \in [400,\, 2000]\,\mathrm{Mpc}$; the  polarization angle was uniformly sampled from $\psi \in [0,\, 2\pi)$. 
We always kept the orbital orientation face-on, i.e., with inclination angle $\iota = 0$. Realistically, one expects the orientation to be random; however, here our focus was not on the effect of varying $\iota$ but on the impact of glitches on detectability. In constructing the GW strain time series, we accounted for the dependence of the detector antenna-pattern functions on the sky position $(\alpha,\, \delta)$ and polarization angle $\psi$. A consequence of distributing the sources uniformly in volume is that their sky positions are isotropically distributed. However, since the simulated signals are injected into real data selected from across O3, the strength of the signal in the detector from two BBHs with identical parameters, except for their times-of-arrival, can differ due to the detector antenna patterns changing with Earth’s rotation. 

Following the simulation of the BBH signals described above, 
{we injected them into a sufficiently long and quiet section of real LIGO-Hanford (H1) detector noise from O3, with no known signal or glitch triggers in GraceDB.}
We then individually computed their 
signal-to-noise ratio (SNR), as defined below,
and retained only those that had SNR$\geq 7$. This SNR cut was performed  
to filter out signals with insufficient strength to ensure reliable signal detection and training. 
The  matched-filter peak SNR on those injections was computed against the true unit-norm template (i.e., with the same intrinsic parameters, such as component masses and spins, as the injection) 
following
\begin{equation}
\text{SNR} \equiv 4\, \mathrm{Re} \int \frac{\tilde{s}(f)\, \hat{\tilde{h}}^*(f)}{S_n(f)}\, df\,,
\end{equation}
where $\hat{\tilde{h}}(f)$ and $\tilde{s}(f)$ are the Fourier transforms of a unit-norm CBC template and the detector strain, respectively, and $S_n(f)$ is the one-sided power spectral density of the detector noise. 
Also, the integration is over the bandwidth of the template.
About 98\% of our initial set of simulated signals survived this cut. The surviving signals were further injected into a large set of different detector noise and glitch segments for training. Each of those simulated GW signals 
is injected randomly into either a quiet noise segment or a transient noise artifact (glitch) segment,
%}, 
with the merger time placed at the center of each segment 
with an additional random shift within $[-2,\, 2]$ seconds. This sky projection, SNR cut, and injection process was repeated for each training epoch, ensuring that the training samples varied between epochs. This exercise was conducted to avoid overfitting to any specific bias patterns.

\begin{table}[ht]
\centering
\caption{Parameters of the simulated binary black hole signals used in our study.}
\label{tab:bbh_parameters}
\begin{tabular}{lcc}
\hline\hline
\textbf{Parameter} & \textbf{Range} & \textbf{Distribution} \\
\hline
Total mass ($M_\text{total}$) & 50--120 $M_\odot$ & Uniform (250 points) \\
Mass ratio ($q = m_2/m_1$) & 0.1--1.0 & Uniform (40 points) \\
Luminosity distance & 400--2000 Mpc & Uniform in volume \\
Polarization angle $(\psi)$ & 0--$2\pi$ rad & Uniform \\
Orbital inclination angle $(\iota)$ & Fixed & 0 (face-on orbit) \\
Sky location & All sky & Isotropic \\
SNR threshold & 7 & -- \\
\hline
\textbf{Simulations surviving SNR cut:} & & \textbf{ $9,820$} \\ 
\hline\hline
\end{tabular}
\end{table}

\subsection{Training and Testing Dataset}

As listed in table~\ref{tab:dataset_composition}, our training dataset included several kinds of real data from O3, namely, 15000 quiet noise segments, 8800 glitch segments and 9820 quiet noise segments with simulated signals injected as well as 
an equal number of glitchy noise segments with simulated signals injected. Whitening was applied to all 30-second segments.
To estimate the PSD, we use Welch's method with 2-second windows over the first 8 seconds of each 30-second segment. The segments are windowed using the Hann window, and the periodograms are median-averaged.
To avoid edge effects, we 
shorten
all segments to 22 seconds length. 
For training,
the final 
samples were created by extracting 5-second windows from each 22-second segment. 
In addition to the BBH signal injection offset in the timeseries, the slice taken for the 5-second window is also offset such that the signal is randomly placed within the central 4 seconds.~\footnote{Two offsets are performed to randomize both the location and separation of transients within the segment if glitches are present. In the glitchy segments, the artifact of interest is always located around the center.}
This procedure enforces time-translational invariance, ensuring that the model remains agnostic to the absolute timing of the signal within the input window. 

To create the test set
we first randomly selected 2,000 quiet noise and 2,000 glitch segments from the 
initial data
pools. 
We then randomly picked 1,000 BBH signals from the original 10,000, after applying the SNR cut.~\footnote{The background segments (quiet or glitchy) chosen for the test set were removed from the training pool, but the BBH
waveforms were not. The latter were left in since the different strain background and BBH waveform projections, for varied sky positions and source distances, produce samples independent from training.} To increase the discrimination challenge in the testing set, we implemented GW signal injections differently for glitches: we did not introduce random time shifts, ensuring that GW signals and glitches were 
overlapping. 
All in all, the test set contains 1000 samples of each scenario. This dataset remained static among different training runs, and was used for all testing of different methods.
For the training set, we utilized 
23,791 segments of quiet detector noise, approximately 9,820 of which with BBH injections,
and approximately 18,629 glitch segments with 9,820 BBH injections.
We had 20 training
epochs 
to establish our
network model.
The training and testing dataset compositions are summarized in table~\ref{tab:dataset_composition}. 

Testing is performed similarly for all search methods employed in this study. Each method was run over each sample in the testing dataset, with the max ``score" of each method around the known event (or non-event) taken as output. Differences arise in testing only between the matched filter and ML methods. ML methods produce only a single output data stream for each sample, whereas matched filtering produces one for each template for each sample.~\footnote{In our testing, a full chi-squared timeseries was generated for each sample for each template, and the SNR timeseries was downweighted with PyCBC's \texttt{newsnr} function~\cite{Nitz_2018}.} 
After all of the templates were correlated with a given sample, the template with the highest match, i.e., the new SNR~\cite{Nitz_2018},
is chosen as the detection event.

\begin{table}[ht]
\centering
\caption{Dataset composition: 
A total of 4000 separate segments were used for testing, evenly divided among the four categories of data types.}
\label{tab:dataset_composition}
\begin{tabular}{lccc}
\hline\hline
\textbf{Data Type} & \textbf{Total Available} & \textbf{Used for Training} & \textbf{Used for Testing} \\
\hline
Quiet Detector Noise & 25,791 & $\sim$15,000 & 1,000 \\
Glitches (Blip \& Scattered Light) & 18,630 & $\sim$8,800 & 1,000 \\
Signal + Quiet Detector Noise & - & $\sim$9,820 & 1,000 \\
Signal + Glitch & - & $\sim$9,820 & 1,000 \\
\hline\hline
\end{tabular}
\end{table}

\subsection{Detection Significance}

Our model outputs {\it logit} values for every signal or noise trigger. %THIS is repeated below: On their own no statistical significance. 
In statistics and machine learning, logit represents the 
natural logarithm of the odds, which is the probability of an event occurring divided by the probability of it not occurring. 
%log-odds of a probability. 
As such, it is the raw, unnormalized output of a model,
particularly in classification tasks, %before it is transformed 
that can be mapped into probabilities. It represents the linear output of a model's final layer, often before applying an activation function like softmax or sigmoid. 
This numerical value indicates the model's confidence or score for a particular class or outcome, and can be converted into a probability measure.

While the model is optimized to maximize the likelihood across the training dataset, the output logits on their own bear little statistical significance on general data. To properly calibrate our model's performance and provide confidence estimates, we build a background distribution (using regular detector noise, with and without glitches) from the GW output logits, 
and perform a 
hypothesis test that a given trigger is sampled from the background distribution and is not a CBC signal.

\begin{figure}[h!]
\centering
\includegraphics[scale=0.5]{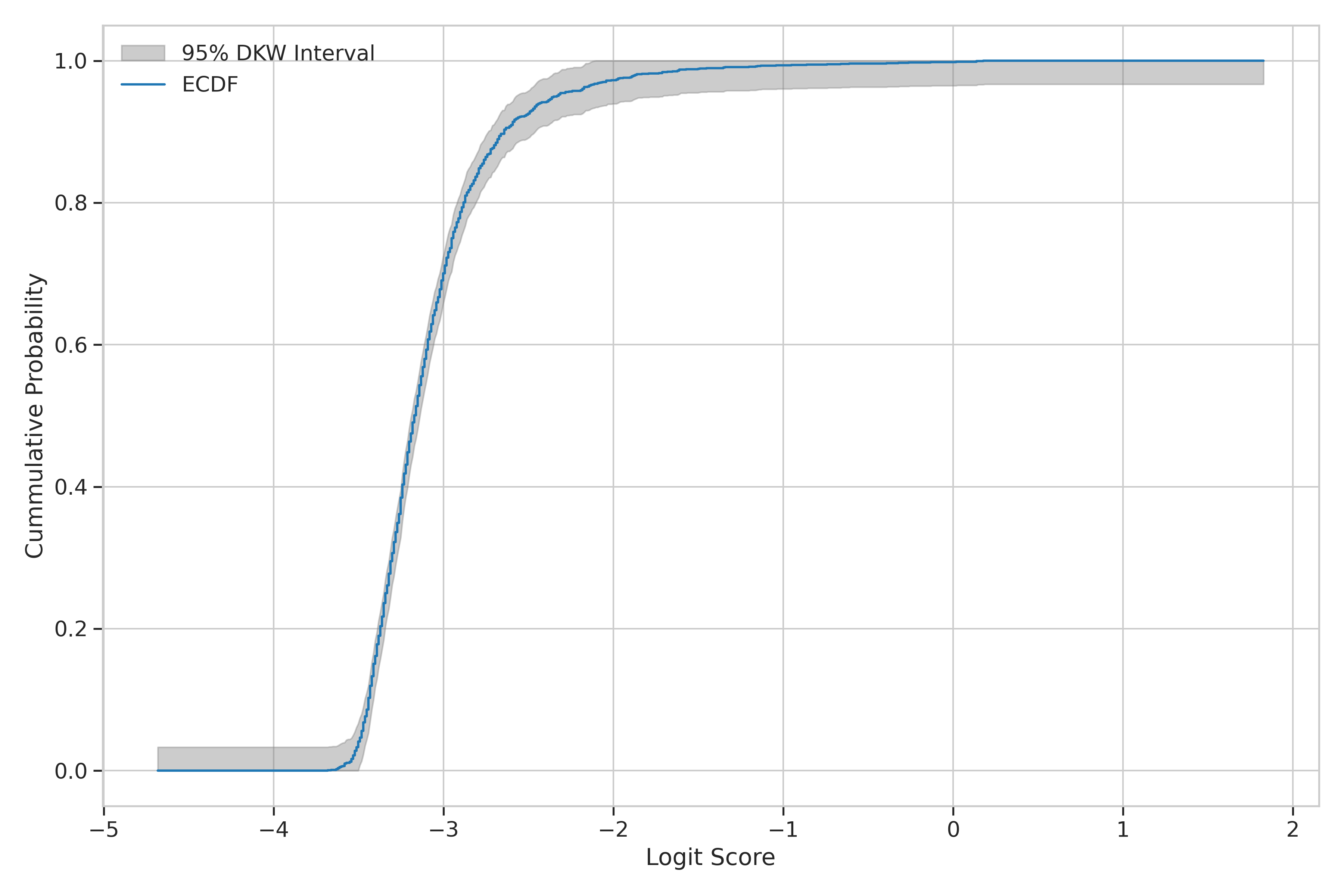}
\caption{Plot of the empirical cumulative distribution function (ECDF) generated from the 2000 background samples in the test set. The grey shaded region is the 95\% confidence interval from the two-sided Dvoretzky-Kiefer-Wolfowitz-Massart (DKWM) ~\cite{massart1990} inequality.}
\label{fig:DKWECDF}
\end{figure}

Due to our finite %test 
background dataset, we built an empirical cumulative distribution function (CDF) -- or ECDF -- to perform this calibration,
as illustrated in Fig.~\ref{fig:DKWECDF}. 
A trigger with a lower logit score is more likely to be noise than signal.  
At any given logit score, the cumulative probability in that figure indicates the fraction of background triggers for which the model returned lower scores.
To quantify the error inherent to our ECDF, we bound it using the Dvoretzky-Kiefer-Wolfowitz-Massart (DKWM) inequality~\cite{massart1990}, as shown by the grey shaded region in Fig.~\ref{fig:DKWECDF}. 
We use this mapping to deduce from logit values the false-alarm probability of triggers in any data type in order to construct the Receiver Operating Characteristic (ROC) curves~\cite{helstrombook} in Sec.~\ref{sec:results}. The ROC curves 
are useful in comparing the performance of the model with that of traditional search pipelines.

\begin{figure}[h!]
\centering
\includegraphics[scale=0.5]{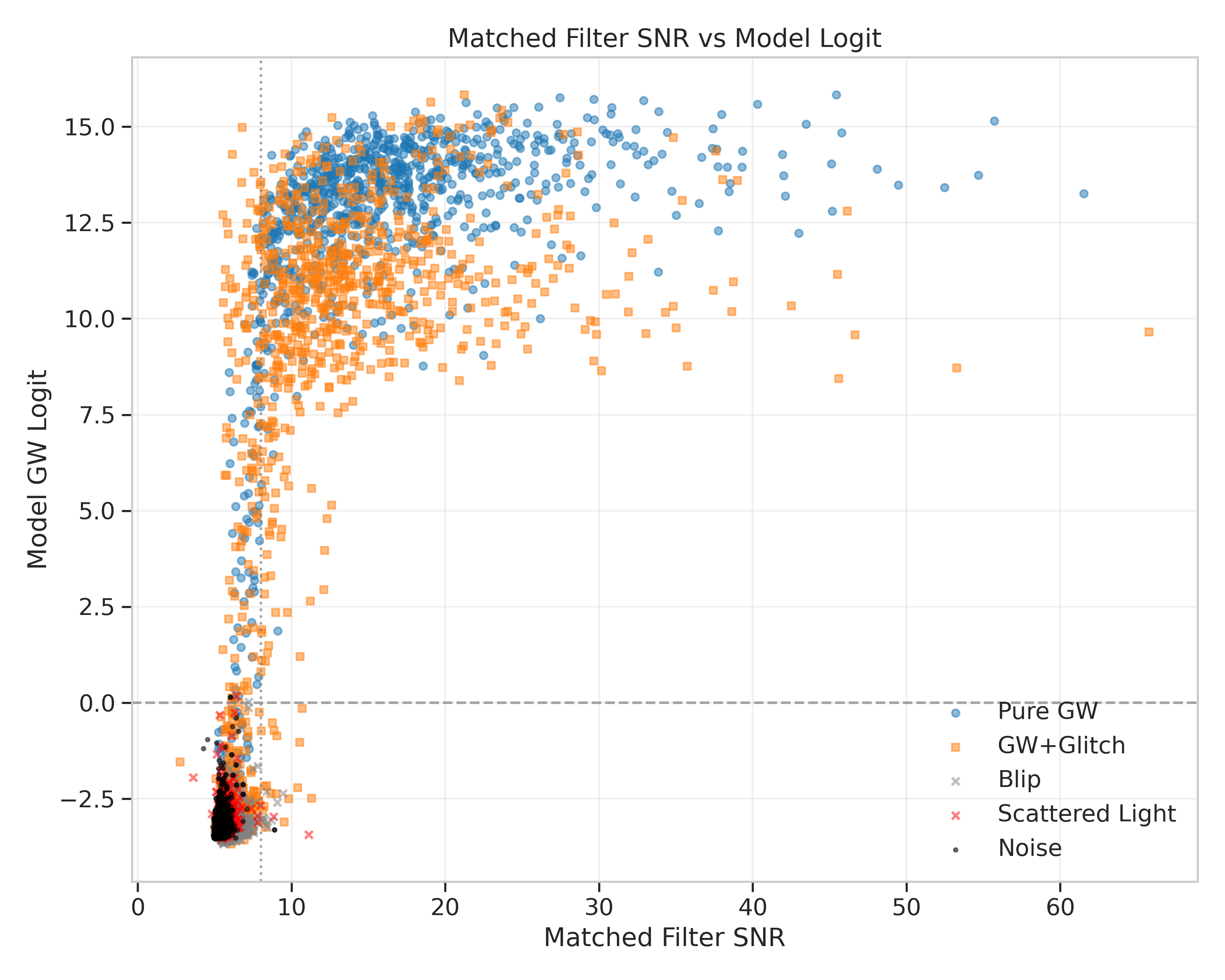}
\caption{Plot of model output logit versus matched filtering SNR over the entire test set.
The vertical dotted line is at an SNR of 8.}
\label{fig:SNRvsLOGIT}
\end{figure}

The first qualitative evidence of overall improvement in the detectability of BBH signals by our Conformer Binary neTwork Search, or \texttt{CoBiTS}, over matched filtering is presented in Fig.~\ref{fig:SNRvsLOGIT}.
There we show the scatter plot of a chi-square-weighted matched-filter SNR -- i.e., the \texttt{newsnr}~\cite{Nitz_2018} -- of every noise and GW trigger along with their logit values obtained from \texttt{CoBiTS}. There are five different categories of triggers represented here, namely, those from: (a) simulated BBH signals added to real, relatively quiet detector data; (b) same as in item ``a'' but when the real data had either a blip or a scattered-light glitch (of either the slow or the fast kind)~\cite{Soni_2021}; (c) real data containing a blip glitch; (d) real data containing a scattered-light glitch;
(c) relatively quiet real detector data. This figure is further divided into four ``quadrants'' by a dotted line at 
\texttt{newsnr}~$= 8$ (labeled as ``Matched Filter SNR" in that figure)
and a dashed line at logit~$= 0$. There is a small fraction of triggers from category ``b'' (GW+Glitch) that have \texttt{newsnr}~$> 8$ but logit $< 0$, i.e., in the bottom-right or ``fourth" quadrant. 
For those signals, the traditional matched-filtering search performed better than \texttt{CoBiTS}.
Comparatively, there are far more  triggers from the first two GW-signal categories, ``a" and ``b", that are in the second (or top-left) quadrant, i.e., have logit~$> 0$ but \texttt{newsnr}~$ < 8$.
This observation suggests that \texttt{CoBiTS}' logit finds more BBH signals with higher significance than \texttt{newsnr}. It is, however, the ROC curve comparisons in Fig.~\ref{fig:ROCs} that verify this.
Triggers in the first (i.e., top-right) quadrant are detected by both \texttt{newsnr} and logit with high enough significance.

\section{Results}
\label{sec:results}

\begin{figure}[h!]
\centering
\includegraphics[width=1\textwidth]{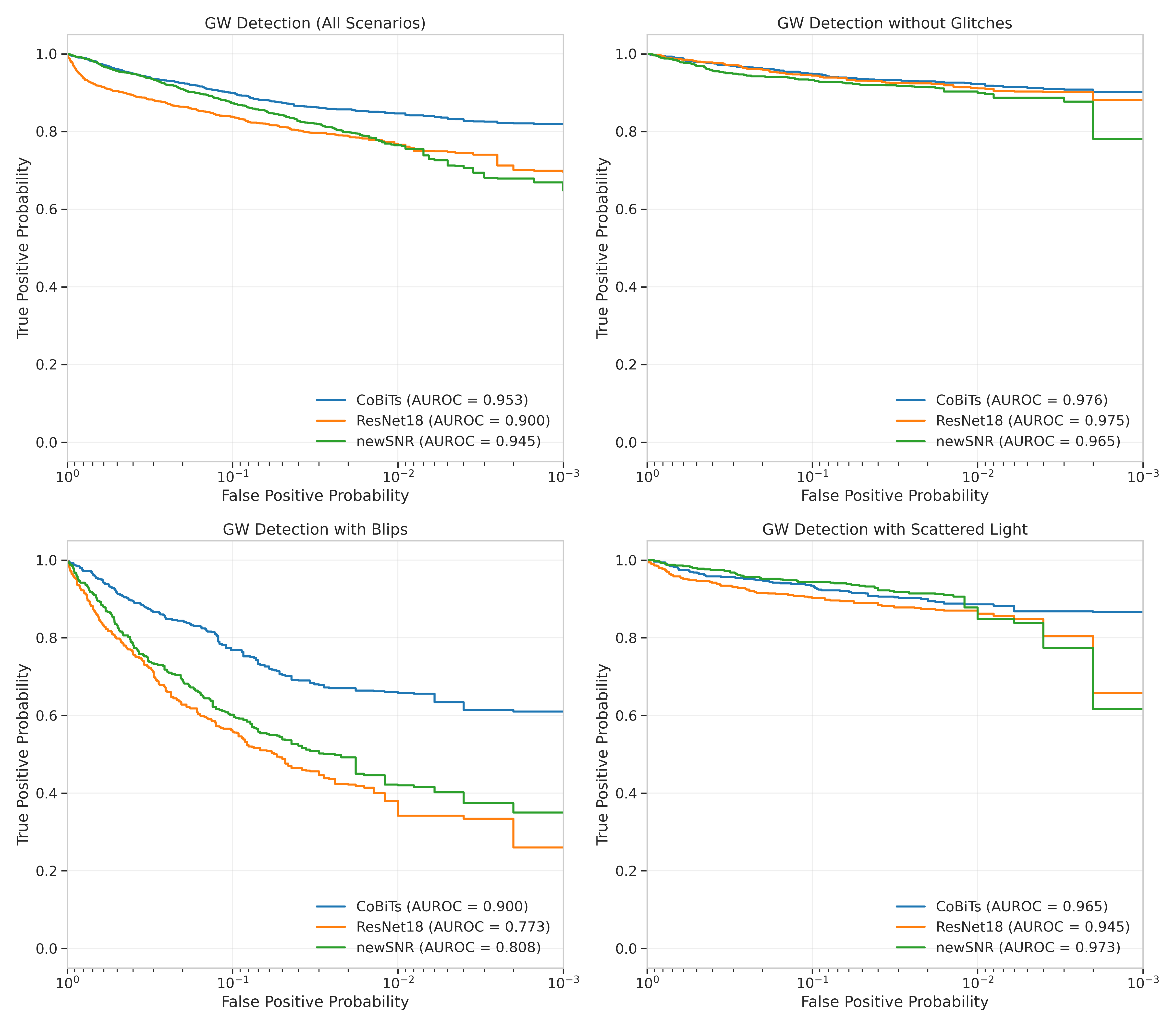}
\caption{
ROC curves of different search methods for BBH signals with and
without glitches in real GW data.
In all cases, \texttt{CoBiTS} outperforms ResNet most likely due to its ability to discriminate based on both local and global attention. All three models perform nearly equally well in the case where the data contains only BBH signals and no glitches. 
}
\label{fig:ROCs}
\end{figure}

Since the second quadrant includes some noise triggers, it is difficult to assess from the scatter plot alone if the apparent better signal recovery by logit is accompanied by a worse false-alarm rate or not, compared to \texttt{newsnr}. 
That assessment is better made by Receiver-Operating Characteristic (ROC) curves, which are shown in Fig.~\ref{fig:ROCs}. There, for three different detection models, i.e., \texttt{CoBiTS}, ResNet, and 
\texttt{newsnr} (which employed a $\chi^2$ weighted 
matched-filtering), 
%(with $\chi^2$) 
we compare how the true positive rate climbs with increase in the false positive rate in various signal-glitch scenarios. 
In all cases, \texttt{CoBiTS} outperforms ResNet most likely due to its ability to discriminate based on both local and global attention. All three models perform nearly equally well in the case where the data contains only BBH signal and no glitches. \texttt{CoBiTS} performs much better than its competitors when a time-segment with a BBH signal also contains a blip glitch. This is the scenario where \texttt{CoBiTS} is expected to be most useful in reducing single-detector based public alerts with low-latency. 
In the absence of glitches, \texttt{CoBiTS} performs nearly as well as  matched-filtering (with $\chi^2$), which is expected since the latter is an optimal search statistic in that noise. However, \texttt{CoBiTS} does perceptively better when glitches are present. Additionally, it is computationally much more economical once it has been trained, prior to deployment.

\begin{figure}[h!]
\centering
\includegraphics[width=1\textwidth]{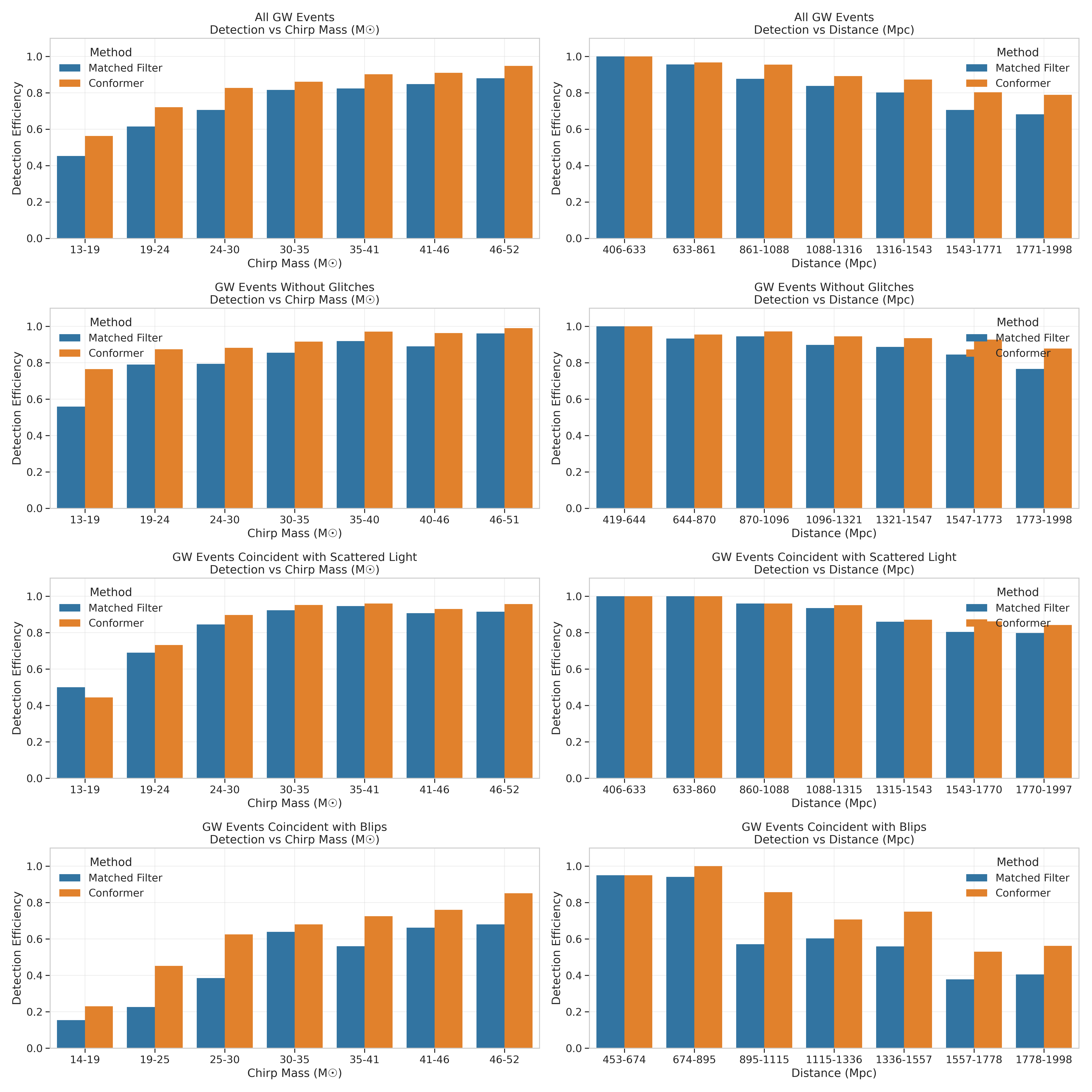}
\caption{Detection efficiencies for various parameter bins -- for \texttt{CoBiTS} and a standard matched-filtering SNR search, which employed the 
$\chi^2$ statistic as described in Ref.~\cite{nitz_distinguishing_2018}. The threshold for detection in this plot is $\mathrm{FAP} < 10^{-3}.$}
\label{fig:deteff}
\end{figure}

\begin{figure}[h!]
\centering
\includegraphics[width=0.6\textwidth]{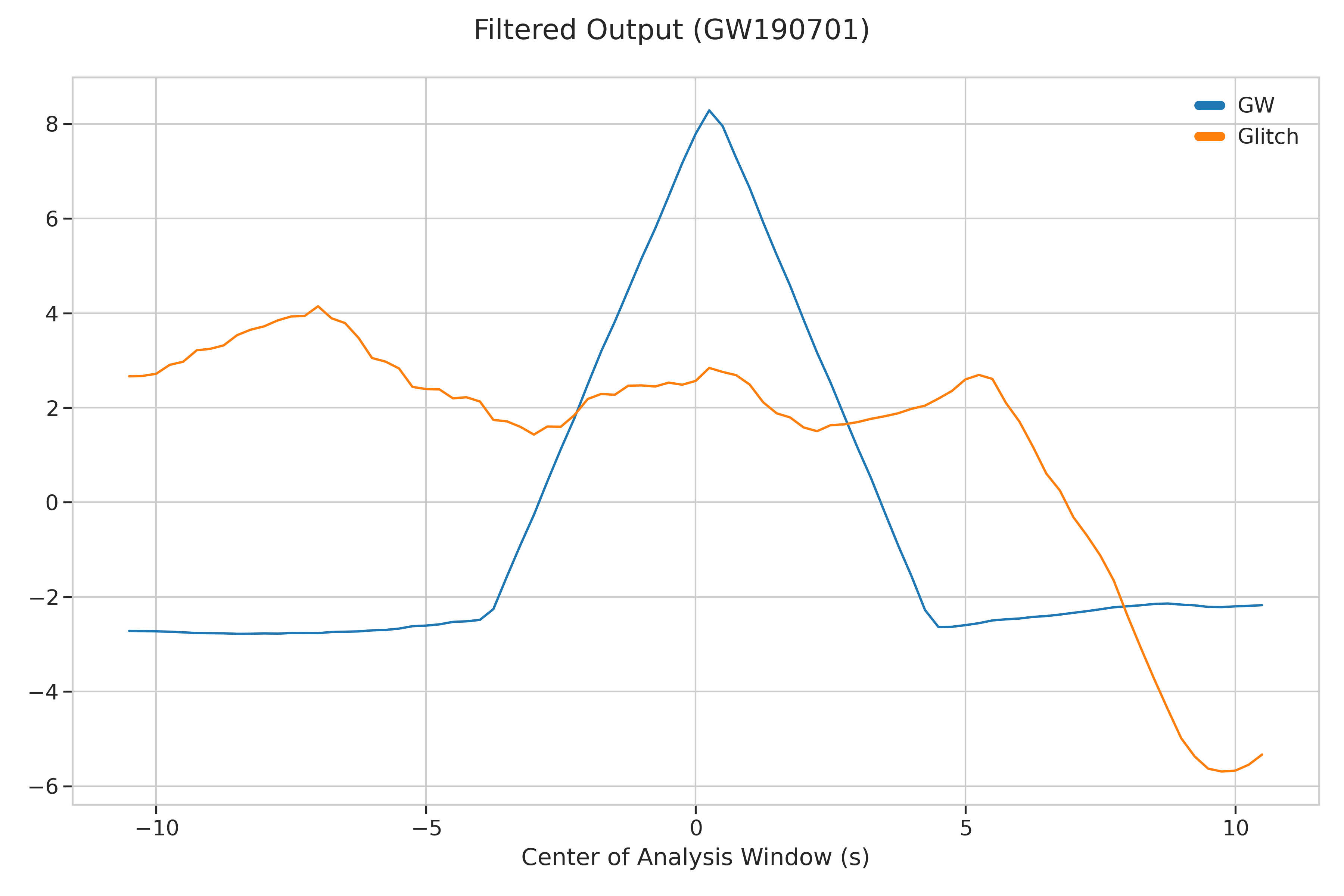}
\caption{Filtered model outputs from GW190701, a relatively high mass BBH merger, in the presence of a scattered light glitch in  data from LIGO-Livingston (L1). \texttt{CoBiTS} was not trained on L1 data.
{The blue trace shows how \texttt{CoBiTS}' logit score (shown on the $y$-axis) rises as its analysis window slides across the strain time-series data containing the (real) BBH event. At later times, that score falls off to values consistent with background detector noise.
The orange trace shows the logit score recognizing the presence that a (slow scattering) glitch, which can last for several seconds. That score is initially high, when the glitch is present, before falling off at later times. 
}}
\label{fig:GW190701}
\end{figure}

\begin{figure}[h!]
\centering
\includegraphics[width=0.6\textwidth]{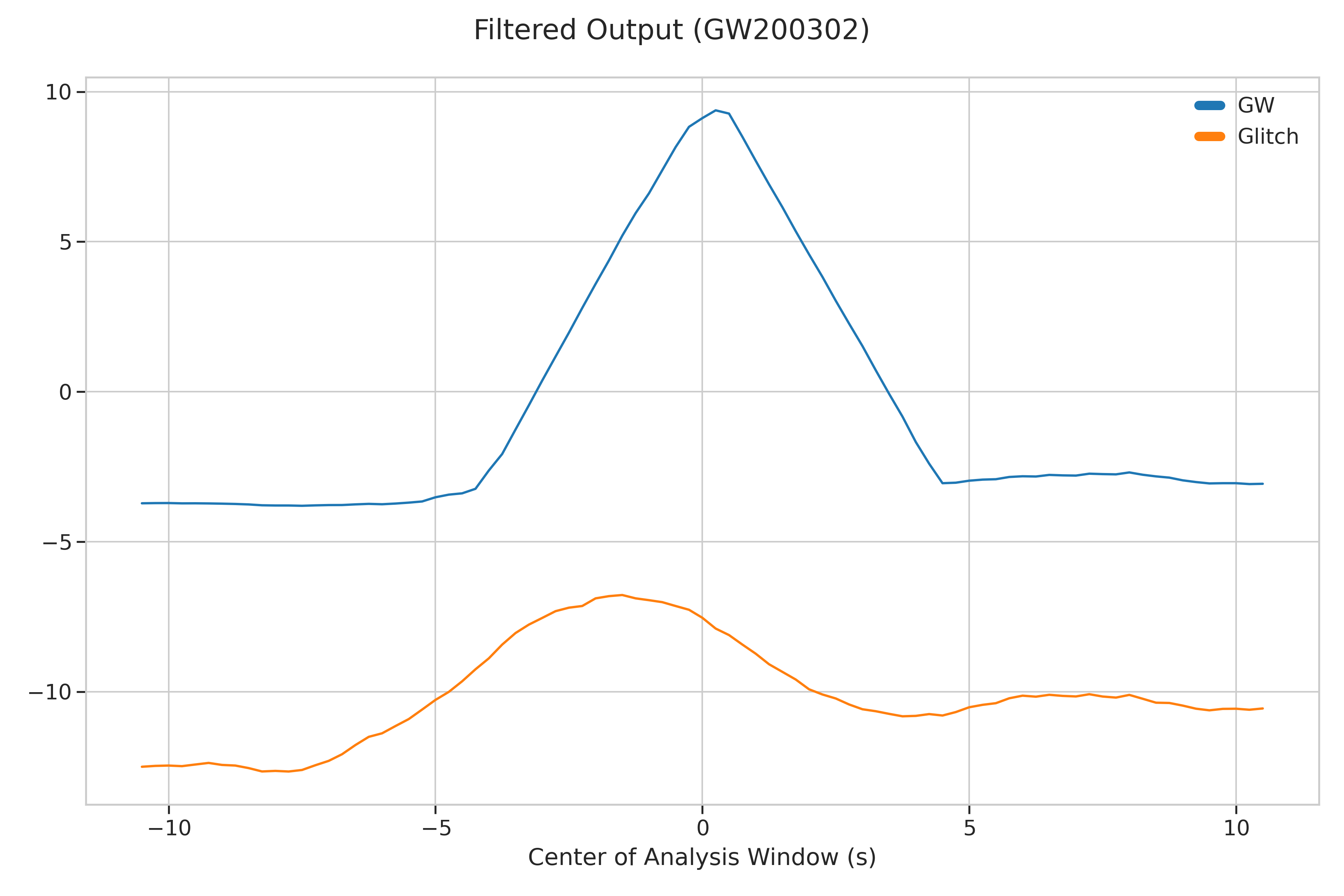}
\caption{Filtered model outputs from GW200302, a relatively high-mass merger event without the presence of glitches. Data was taken from H1.}
\label{fig:GW200302}
\end{figure}

In Fig.~\ref{fig:deteff}, we compare the detection efficiencies of our Conformer model \texttt{CoBiTS} with the traditional method for various scenarios, in chirp-mass and distance bins. \texttt{CoBiTS} performed better in all bins and all cases, except for a few distance bins where the performances are comparable.
In order to interpret better how \texttt{CoBiTS}' perception about the presence or absence of signal / glitch evolves as a timeseries snippet is read in, we plot in 
Figs. ~\ref{fig:GW190701} and ~\ref{fig:GW200302}
the time-evolution of its assessed probability for two different scenarios for what the data contain: (a) Only a BBH signal; (b) Only a glitch (blip or scattered light).
What is not shown is the case where the data contain just quiet noise, without any known BBH signal or glitch logged in GraceDB~\cite{gracedb}. For this last case, both the signal and glitch scores turn out to be quiet low, as illustrated by the orange trace in Fig.~\ref{fig:GW200302} before -5s and after 5s.
For this last scenario, the probability traces for the hypotheses that the data-snippet contains a BBH signal or a glitch remain  close to zero. 
Of the 1000 quiet noise samples in the test set, for 99\% the logit score lies below -1.349.

\begin{figure}[h!]
    \centering
    \includegraphics[scale=0.5]{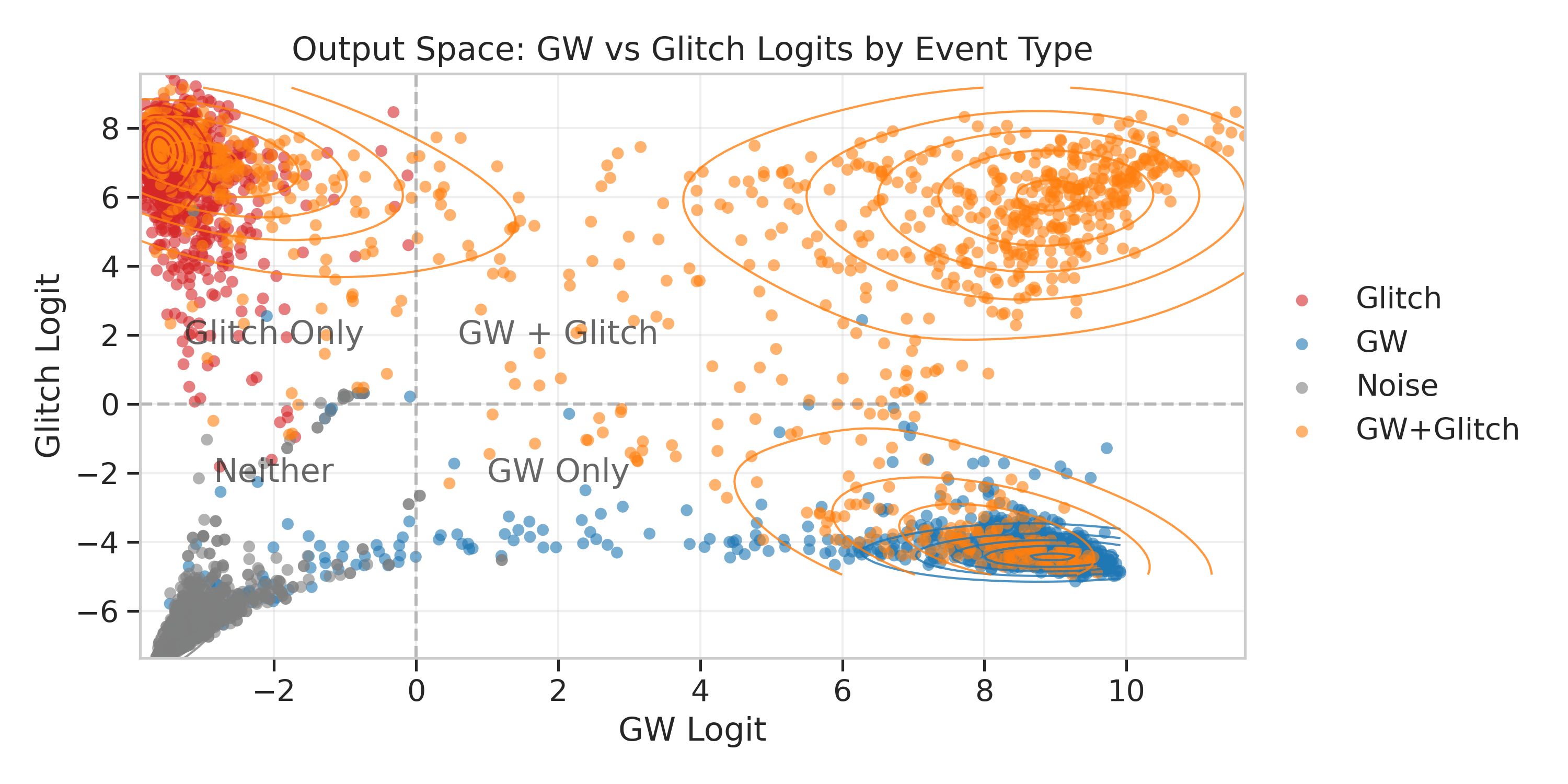}
    \caption{Scatter plot and KDE over the model output space, generated from evaluation on the testing dataset. The color of each point indicates the true value of the corresponding sample.}
    \label{fig:outputscatter}
\end{figure}

In Fig.~\ref{fig:outputscatter} we provide further insights into the capabilities of \texttt{CoBiTS} by making a scatter plot of the aforementioned four scenarios of data with / without BBH signals and glitches, and quiet data.
To interpret this figure, imagine dividing it into four quadrants where points representing data that contain louder and clearer BBH signal (glitch) features will be found more to the bottom-right (top-left). Data that are clearly discerned to contain both BBH signals and glitches are found in the top-left and quiet noise in the bottom-left. 
A trail of BBH signal-only (glitch-only) triggers, with increasing strength, bridge the divide between the noise-only cluster and the BBH-signal (glitch) cluster, as one would expect. Loud enough BBH signals, with or without overlapping glitches, are all clustered on the right of the figure. Signals to the right of the vertical line at GW logit $=0$ have a FAP of less than 0.002. 
The colored contours are obtained using kernel density estimation (KDE) and indicate where \texttt{CoBiTS} assesses the classification significance to be the highest (e.g., the central yellow contour shows the most significant logit values for a data snippet to have both a GW and a glitch. The wider contours indicate how slowly or rapidly that classification significance reduces over the GW logit and the Glitch logit.

\subsection{Performance on Real Events}

{Since our test set consists purely of simulated signals injected into real data (here shown only for H1 segments), it is important to verify the model's performance over real, cataloged events. Testing was performed over a litany of different events, both within and outside of the trained parameter space of the model. The results of these tests show a strong generalization of \texttt{CoBiTS}. }
{We highlight a few of those examples here.}

\subsubsection{GW190701 \& GW200302}

\begin{figure}[h!]
\centering
\includegraphics[width=0.9\textwidth]{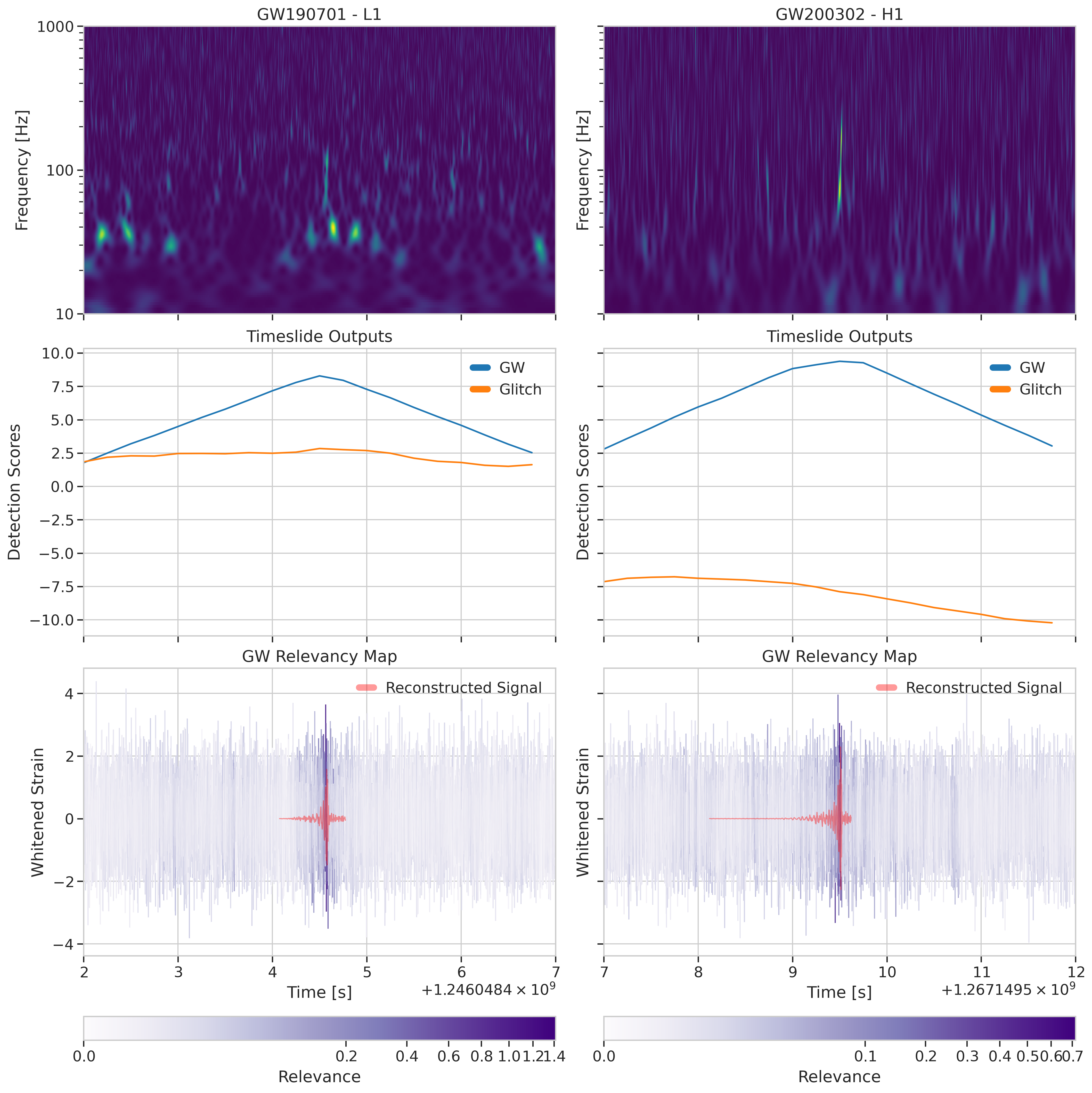}
\caption{Left and right columns correspond to GW190701 and GW200302, respectively. 
Top row: Omegascan of each event.
Middle row: \texttt{CoBiTS} output for the corresponding event.
Bottom row: \texttt{CoBiTS} relevance map. The colored background shows the whitened strain, and the red curve is the whitened, reconstructed signal. The background color indicates the local contribution of each time–frequency point to the \texttt{CoBiTS} GW output logit.}
\label{fig:3stackrealevent}
\end{figure}

These events showcase \texttt{CoBiTS}' performance and generalization on cataloged events within the trained merger parameter space. In Fig.~\ref{fig:3stackrealevent}, left column, the real BBH event GW190701 demonstrates this well; this BBH signal was concurrent with a scattered light artifact in the Livingston detector. Although trained exclusively on data from H1, \texttt{CoBiTS} correctly identifies this simultaneous occurrence of a real signal and a glitch in data from another detector (namely, L1). 
For comparison, Fig.~\ref{fig:3stackrealevent}, right column, shows \texttt{CoBiTS}' signal and glitch logit scores for GW200302 
-- a BBH event lying within the same trained signal parameter space -- but in H1 data not known to contain any glitches. 
While thorough testing of generalization to L1 and other observing runs will be demonstrated in future work, empirical evidence from multiple cases so far shows strong generalization to both other observing runs and detectors.

\subsubsection{Events \& Glitches outside of the training space}

\begin{figure}[h!]
\centering
\includegraphics[scale=0.5]{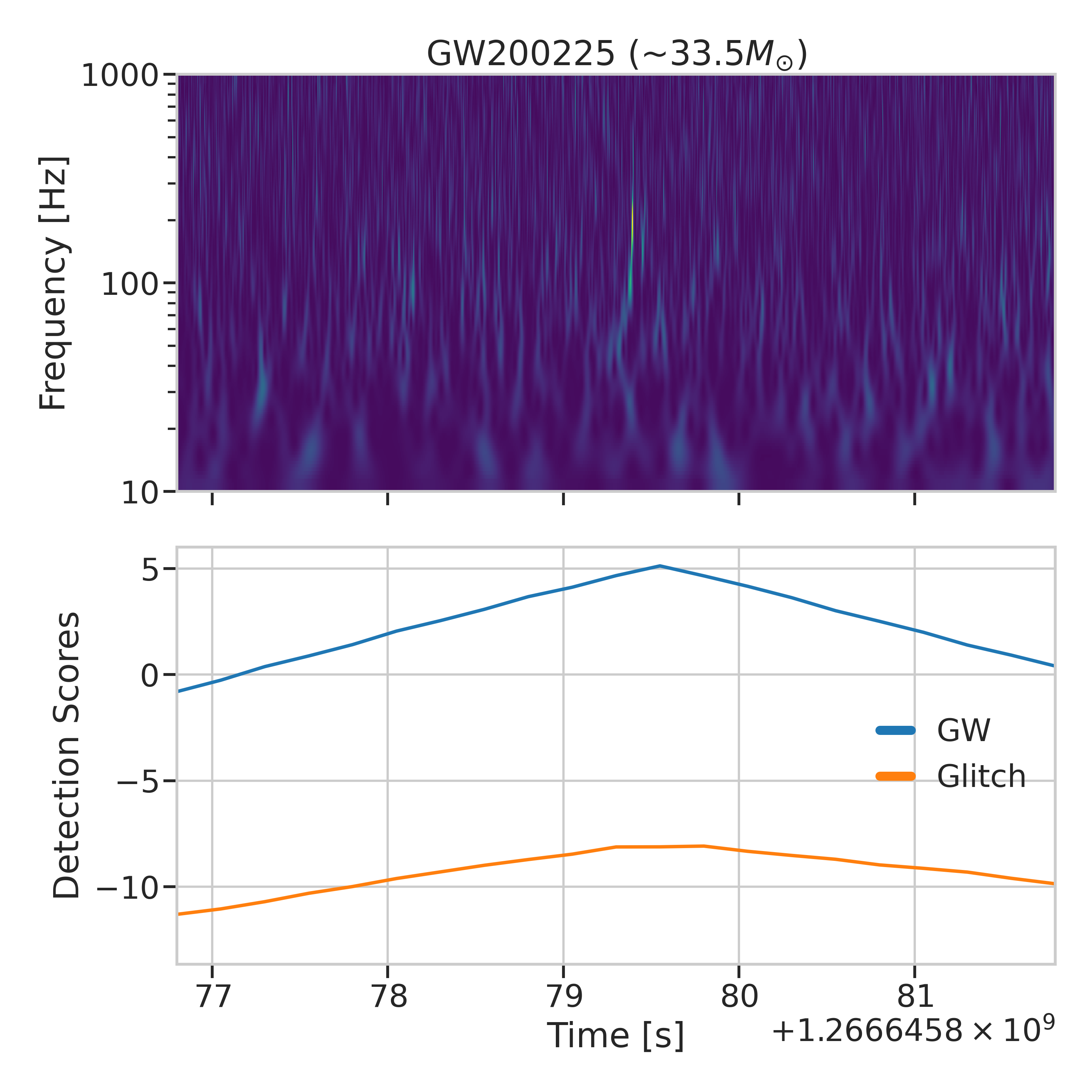}
\caption{Model analysis over a BBH event with the total mass outside of the training space. \texttt{CoBiTS} is only trained to identify BBH with total mass between 50 and 120 $M_{\odot}$, yet it extrapolates and correctly identifies GW200225. The strain analyzed here was taken from H1.}
\label{fig:GW200225Test}
\end{figure}

\begin{figure}[h!]
\centering
\includegraphics[scale=0.5]{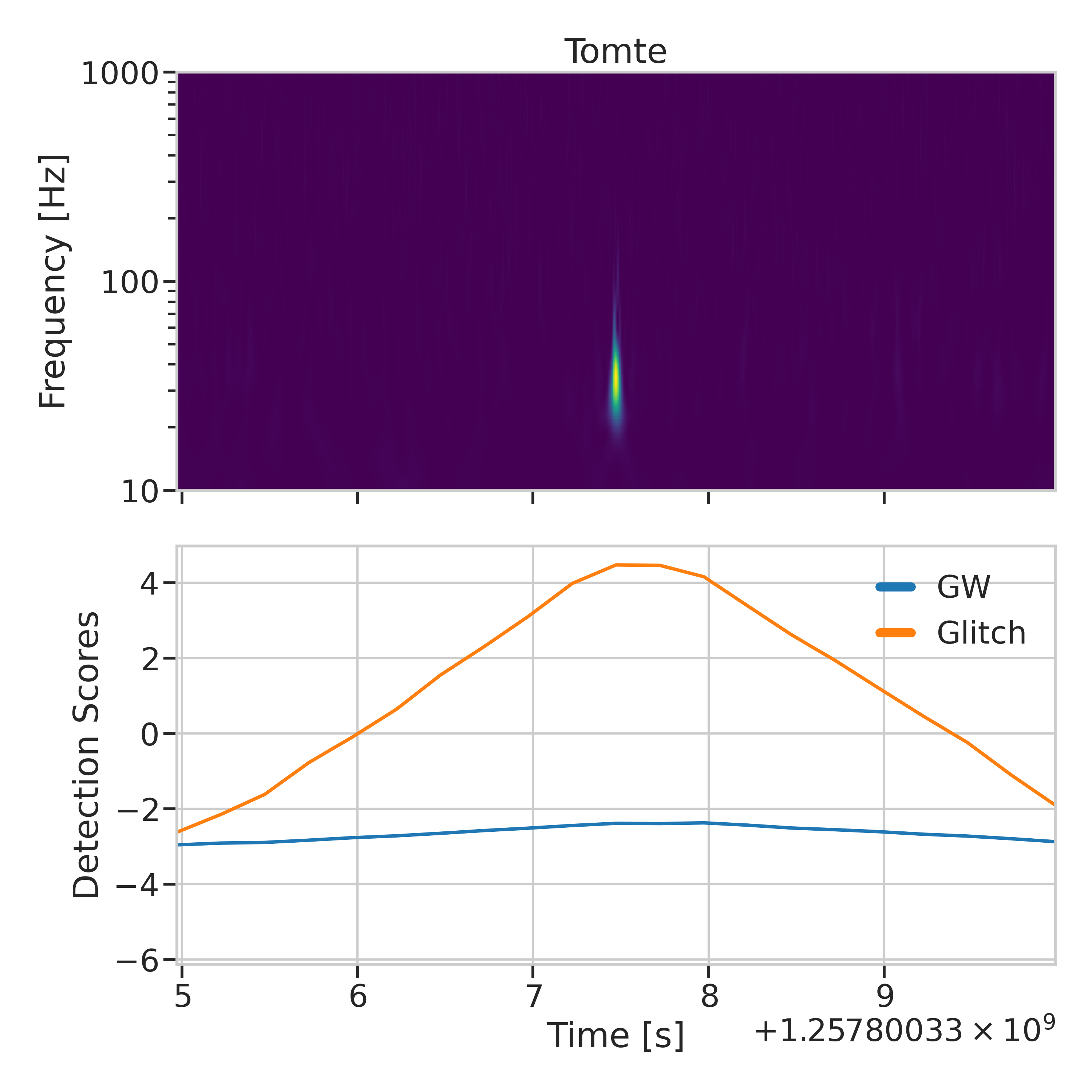}
\caption{Model analysis over a glitch type outside of the training space. \texttt{CoBiTS} here correctly identifies Tomte as a glitch, without ever seeing it in training.}
\label{fig:TomteTest}
\end{figure}

\begin{figure}[h!]
\centering
\includegraphics[scale=0.5]{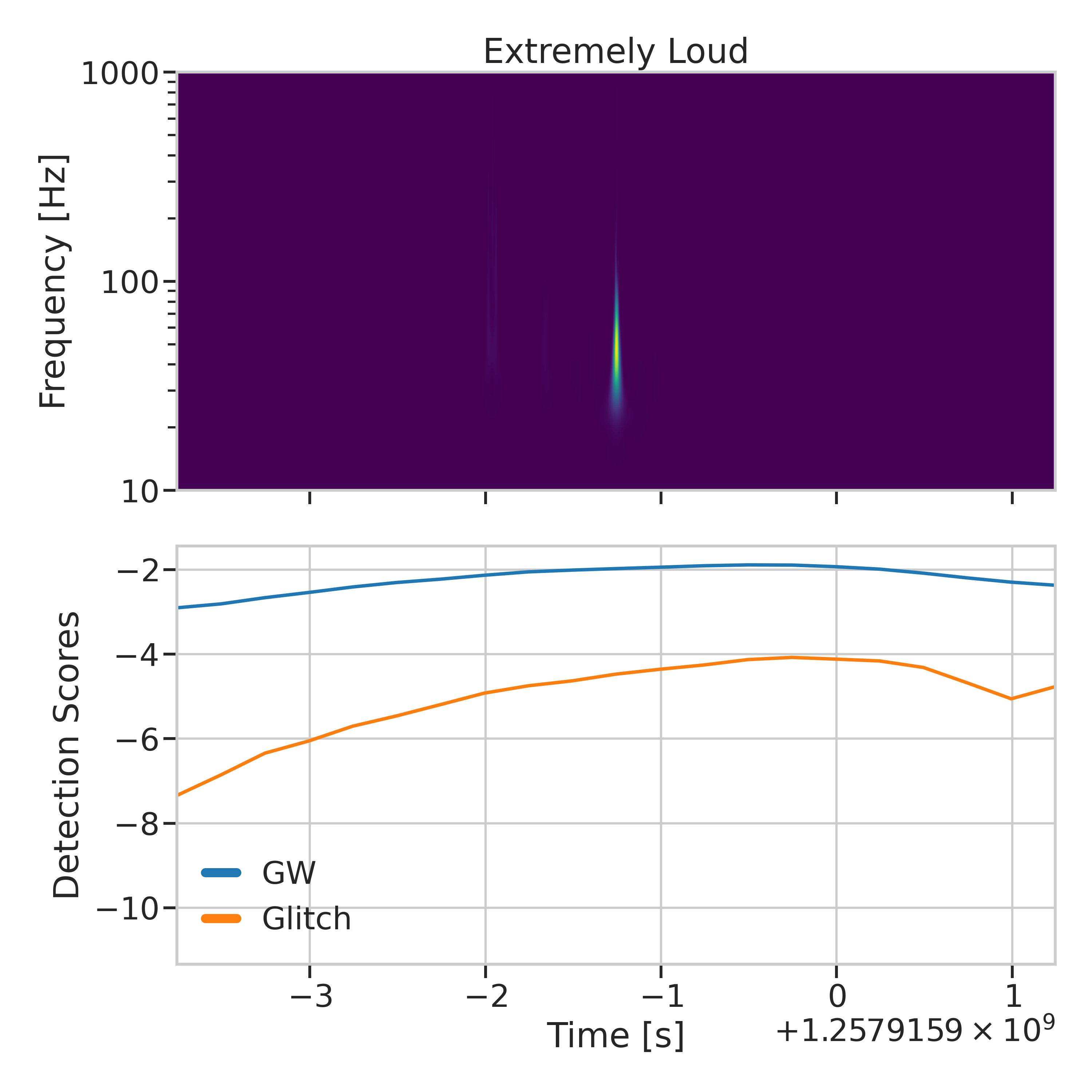}
\caption{Model analysis over a glitch type outside of the training space. \texttt{CoBiTS} here does not identify the Extremely Loud glitch with a large enough glitch score. However, it importantly also disregards it as a potential CBC.}
\label{fig:XLoudTest}
\end{figure}

{With \texttt{CoBiTS} trained on only a subset of the space of signals and glitches, its results when analyzing events outside of the training space are important to assess. This regime is where the generalization of \texttt{CoBiTS} appears the strongest, as it can effectively identify both glitches and signals outside of what it was trained on. Consider, e.g., GW200225, with an estimated total mass of $33.5 \mathrm{M_\odot}$. \texttt{CoBiTS}, trained on signals between 50 and 120 $\mathrm{M_\odot}$, correctly identifies this as a BBH event. Fig.~\ref{fig:GW200225Test} shows the model's results. Figs.~\ref{fig:TomteTest} and~\ref{fig:XLoudTest} show \texttt{CoBiTS}' outputs over a Tomte and Extremely Loud glitch, respectively. The Tomte is correctly identified as a glitch, while the Extremely Loud artifact is not. However, importantly, \texttt{CoBiTS} doesn not assign a GW signal score higher than approximately -2 for either of these glitches.}

\subsection{Explainability}

Explainability in machine learning is the ability to understand how and why an ML model makes a specific decision or prediction. It 
is a still evolving, 
unresolved question that
aims to solve the "black box" problem, thereby making complex algorithms more transparent by providing explanations for their internal processes and outputs that are comprehensible to
humans. 
This broadly applies to the entire paradigm of ML. Instead of designing a purpose-built workflow, say the matched filter, ML is data-driven, with precise explainability that is challenging to determine. This often results in models getting labeled as the aforementioned black boxes, where the details of the operations between input and output are intractable. While this label has some merit, it is a misrepresentation of ML. Models are not ``black boxes", just obfuscated boxes with a well-understood structure. Much work goes into architecting and training a model, which informs the general operation. However, models -- particularly deep-learning models -- have many weights (our model has on the order of $10^8$), and understanding how these models work with millions, if not billions, of tunable parameters is daunting. Fortunately, work has gone into answering what happens between input and output in deep-learning models. 

\begin{figure}[h!]
\centering
\includegraphics[width=1.2\textwidth]{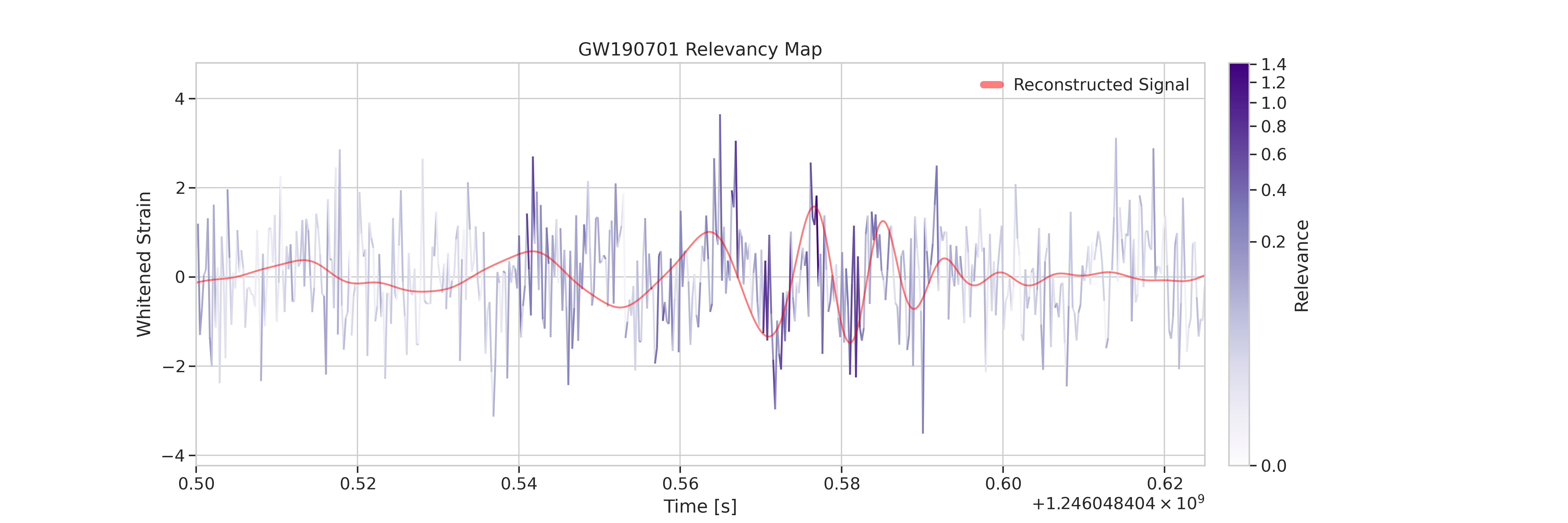}
\caption{Zoomed in version of the bottom plot of~\ref{fig:3stackrealevent}. The red trace is the reconstructed waveform, and the coloring of the strain represents how relevant each point was to determining the output of the model.}
\label{fig:rel_zoom_190701}
\end{figure}

To provide transparency on what parts of the data weigh most in \texttt{CoBiTS}'s decision on the presence of  distibguishing features, we  utilize Integrate Gradients (IG)~\cite{sundararajan2017axiomaticattributiondeepnetworks}. 
IG captures the attribution of each input sample (i.e., at each time step) in determining each output (i.e., if the data snippet contains a GW, glitch, etc.).
Fig.~\ref{fig:rel_zoom_190701}
shows this attribution -- or 
saliency -- in a plot 
of a single forward pass of \texttt{CoBiTS} over the binary black hole signal GW190701. 
A saliency map is an image that highlights  the region in the data that is most relevant to a machine learning model for arriving at its decision.
The purple background trace is the real, whitened input, colored by each sample element's attribution, or relevance, to the GW output, and the overlaid red curve shows  the reconstructed signal. In this particular case, the deeper the purple hue the more relevant those data points were to \texttt{CoBiTS}' assessment that this time series is consistent with a BBH signal. 
Figure~\ref{fig:3stackrealevent} bottom 2 plots
show the same attribution plot for 2 events we test on, but zoomed out and stacked with the Q-scan and model's outputs over the segment~\cite{chatterji_multiresolution_2004}.

\section{Conclusions}
\label{sec:conclusions}

In this work, we introduced one of the first ML models that inputs time-series strain data from a single detector
and outputs the significance of a transient in it as a BBH signal, a glitch or a BBH signal overlapping with a glitch. While it holds promise for functioning as a low-latency multi-detector CBC search pipeline, here we are launching it as a low-latency follow-up tool for vetting of  triggers arising in online CBC searches, e.g., as
uploaded to GraceDB. \texttt{CoBiTS} jobs can be readily launched by   the Rapid Response Team (RRT)~\cite{Chaudhary:2023vec}
for vetting, just like time-frequency spectrograms of trigger data that are constructed for a similar follow-up purpose. \texttt{CoBiTS} can reduce false retractions that may arise due to the presence of a glitch in the neighborhood of or overlapping with a genuine BBH signal, even when it is initially detected in only one detector.

Among follow-up ML tools, \texttt{CoBiTS} has some similarity with GSpyNetTree in that they are both ML based and run in the follow-up mode. However, unlike the latter, which is CNN-based and runs on images of time-frequency spectrograms, \texttt{CoBiTS} runs on strain data itself. Also, as far as we can tell, GSpyNetTree does not evaluate  significances of the same data snippet as containing both a BBH signal and a glitch. GSpyNetTree does classify transients into more categories of glitches and GW signals. The next version of \texttt{CoBiTS} will include better tuning for improved performance against a broader set of glitch-types.

In its current implementation, \texttt{CoBiTS} performs particularly well in  distinguishing BBH signals from blips and scattered light glitches, even when a signal overlaps with a  glitch. 
However, as the ROC  comparisons show, \texttt{CoBiTS} does not perform uniformly better than the matched-filter search, across the FAP range studied here, when all GW and glitch cases are combined. Indeed, in some regions of the FAP, the matched-filter search or ResNet do better. 
Nevertheless, when the areas under the ROC curves (i.e., AUROCs) are compared then \texttt{CoBiTS} usually does better, or at least as good as one of the alternatives. \texttt{CoBiTS}' main weakness is in the lowest chirp-mass bin studied, with ${\cal M} \in [13,19]M_\odot$, when compared to scattered light glitches, as noticeable in the performance histogram comparisons. It is possible to optimize \texttt{CoBiTS} further to address this weakness but here we erred on the side of not overfitting. We did so to safeguard against vetoing 
exceptional BBH events.
More events would presumably increase the chance of netting some exceptional CBCs, such as strongly spinning CBCs or those exhibiting spin flips or transitional precession, intermediate mass black holes (in binaries) and high mass-ratio CBCs.

\section*{Acknowledgments}
\label{sec:acknowledgments}

We would like to thank Tomislav Andric for carefully reading the masnuscript and making helpful suggestions. 
The computational work reported here was supported by the HPC cluster, Kamiak, at WSU. Support from the NSF under Grant PHY-2309352 is also acknowledged.
This research has made use of data or software obtained from the Gravitational Wave Open Science Center (gwosc.org), a service of the LIGO Scientific Collaboration, the Virgo Collaboration, and KAGRA. This material is based upon work supported by NSF's LIGO Laboratory which is a major facility fully funded by the National Science Foundation, as well as the Science and Technology Facilities Council (STFC) of the United Kingdom, the Max-Planck-Society (MPS), and the State of Niedersachsen/Germany for support of the construction of Advanced LIGO and construction and operation of the GEO600 detector. Additional support for Advanced LIGO was provided by the Australian Research Council. Virgo is funded, through the European Gravitational Observatory (EGO), by the French Centre National de Recherche Scientifique (CNRS), the Italian Istituto Nazionale di Fisica Nucleare (INFN) and the Dutch Nikhef, with contributions by institutions from Belgium, Germany, Greece, Hungary, Ireland, Japan, Monaco, Poland, Portugal, Spain. KAGRA is supported by the Ministry of Education, Culture, Sports, Science and Technology (MEXT), Japan Society for the Promotion of Science (JSPS) in Japan; the National Research Foundation (NRF) and Ministry of Science and ICT (MSIT) in Korea; Academia Sinica (AS) and National Science and Technology Council (NSTC) in Taiwan.

\appendix

\section{The Conformer Model Architecture}
\label{subsec:model_arch}

\subsection{Multi-head Feature Extractor}

%\sukanta
{A typical neural network comprises a set of layers, which are a collection of interconnected nodes or {\it neurons}, that function together to process information. These layers are organized into three primary types: input, hidden, and output. The input layer receives the raw data, the hidden layers perform computations and transformations, and the output layer produces the final results.}
As shown in Fig.~\ref{fig:NNworflow}, 
the Multi-Head Feature Extractor (MHFA) is the first module applied to the input data. It acts as a denoising and down-sampling operation, simultaneously embedding the tokens into a higher-dimensional space and reducing the overall sequence length. Through multiple parallel multi-level temporal convolutional modules, the input $[B, 1, T]$ is transformed to $[B, d_{\mathrm{model}}, S]$. There are $N$ heads, where each head is an independent multi-level convolutional module. Each head embeds the input into an embedding dimension of size $\frac{d_{\mathrm{model}}}{N}$ through various learned kernels. Each convolutional layer is followed by layer normalization~\cite{leiba2016layernormalization} and a GELU activation~\cite{hendrycks2023gaussianerrorlinearunits}. Each head may have different kernel sizes (and thus receptive fields), but the stride of each layer is kept constant across all heads to ensure a consistent reduction in sequence length. The product of the strides determines the overall reduction in sequence length from $T$ to $S$. After passing through all $N$ heads, the outputs are concatenated along the embedding (channel) dimension to aggregate information across various receptive fields.

This module can be viewed as a learned filter bank with small templates, scanning the input sequence for local informative features and producing a richer, shorter representation for the Conformer. It is designed to (1) mitigate noise by extracting local, salient features and (2) reduce sequence length to improve computational efficiency. Transformers --and self-attention mechanisms in general --are highly susceptible to noise. This comes directly from their global context, weighing every token in the input sequence. Filtering the timeseries with convolutional layers yields a feature-dense representation of the original input stream for the Conformer. Moreover, long inputs are costly because self-attention scales as $O(S^2)$ in sequence length. Thus, using strided convolutions to downsample the input reduces the model's sequence length and computational cost.

\subsection{Conformer}

The Conformer encoder architecture follows the design introduced by Gulati et al.~\cite{gulati2020conformer}. Attention-based architectures have gained popularity due to their ability to model global context across the entire input sequence. We incorporated self-attention into the model architecture to achieve higher performance on our dataset, and the Conformer-based model emerged as our best-performing approach. Specifically, the Conformer merges the global context of Transformers with the local context of CNNs, thereby achieving superior performance over either alone. Our encoder consists of multiple stacked Conformer blocks.
We trained models with and without a positional encoding scheme and found no difference in model performance. This is likely due to temporal positional information being implicitly captured by the convolutional modules~\footnote{The attention mechanism, on its own, has no sense of the order or position. This is directly resultant from its global context, viewing all tokens the same. Thus, to enable Transformer-style models to understand the relative positioning of tokens, an encoding/embedding schema must be used. Since our input is of a consistent shape, we simply add a learned embedding prior to the encoder layers.}.

\subsection{Final Classification}

The final layer in the model involves pooling across the temporal dimension, followed by a reduction of $d_{\mathrm{model}}$ to two output logits. Rather than using max or average pooling, we employ a learned pooling scheme. A weighted sum is performed across the
sequence, with the weights determined by a small dense sub-network operating along the embedding dimension. After the pooling operation, the data is of shape $[B, d_{\mathrm{model}}]$, and another small fully-connected module reduces the dimension from $d_{\mathrm{model}}$ to 2. GELU activation~\cite{hendrycks2023gaussianerrorlinearunits}
is used again.
{The only place an alternative scheme is used is in the temporal pooling sub-network, where a tanh function is applied as activation before the last fully-connected module. Learned pooling is a type of adaptive pooling. Adaptive pooling refers
to pooling layers that adjust their behavior based on the input data. 
In learned pooling, the pooling operation itself is learned during
training. Traditional pooling methods like max or average pooling use 
fixed kernel sizes and stride. Whereas adaptive pooling, including
learned pooling, dynamically determines these parameters.
Specifically in the case of learned pooling, the pooling operation
itself determines these parameters to optimize performance for the specific task and data.}

\subsection{Training}

All of our various model iterations were trained on the Kamiak cluster at WSU with a single Nvidia H100 GPU. Using a batch size of 32, over 30 epochs, and in 32-bit precision, it takes approximately 12 hours to train a model from initialization.

Due to our dynamic dataset, with an effectively unbounded training corpus, we forego the use of a typical validation set. We track various performance metrics over each batch---such as loss, precision, recall, and F1 score---and halt trials if there is no overall decrease in the loss over five epochs.

\subsubsection{The Objective function and Optimization}

The objective function employed is the binary cross-entropy (BCE) loss, as implemented in PyTorch. Each output logit $z_i$ is transformed into probabilities via the sigmoid function: $p_i = \sigma ( z_i ).$ Each dataset category label (i.e., GW, glitch, etc.) is modeled as an independent Bernoulli trial, resulting in the following joint negative log-likelihood per batch sample:
\begin{equation}
\label{loglikelihood}
-\ln\mathcal{L} = -\sum_i y_i\ln p_i + (1-y_i)\ln(1-p_i) = \sum_i\mathrm{BCE}(p_i)\,.
\end{equation}
Thus, minimizing the BCE loss corresponds to performing maximum-likelihood estimation for the Bernoulli-distributed binary classification outputs. The metric minimized in practice is the above BCE loss averaged across all samples in a batch.

We chose AdamW~\cite{loshchilov_decoupled_2019} as our optimizer for fast convergence and effective regularization (to discourage overtraining and promote generalization). Implemented within PyTorch, the optimizer parameters remain at their default values except for the learning rate and weight decay. During training, we use the OneCycleLR scheduler~\cite{smith2018onecycle}, also provided by PyTorch, rather than a constant learning rate. We found optimal convergence using two warm-up epochs and a maximum learning rate of $10^{-4}$. To stabilize training, we clip gradients at 1.0, apply dropout~\cite{hinton2012} of 0.15 in every module for regularization, and enforce a significant weight decay of $10^{-2}$.

\bibliographystyle{unsrt}
\bibliography{references.bib}

\end{document}